\documentclass[aps,prl,reprint,superscriptaddress,nofootinbib]{revtex4-2}

\usepackage{amsmath,amssymb,bm}
\usepackage{booktabs}
\usepackage{graphicx}
\usepackage{microtype}
\usepackage{xcolor}

\newcommand{\bM}{\ensuremath{\beta_{M0}}}
\newcommand{\bMiv}{\ensuremath{\beta_{M0}^{(M1_v)}}}
\newcommand{\Lnp}{\ensuremath{L_{np}}}
\newcommand{\Lt}{\ensuremath{\widetilde L_{np}}}
\newcommand{\dd}{\mathrm{d}}
\newcommand{\JW}[1]{\textcolor{black}{#1}}

\begin{document}

\title{Percent-Level Prediction of the Deuteron Magnetic Polarizability}

\author{Rishi Cherukuri}
\email{rcheruku@terpmail.umd.edu}
\affiliation{Department of Physics, University of Maryland, College Park, Maryland 20742}

\author{Jiunn-Wei Chen}
\email{jwc@phys.ntu.edu.tw, corresponding author}
\affiliation{Department of Physics and Center for Theoretical Physics, National Taiwan University, Taipei 10617}
\affiliation{Physics Division, National Center for Theoretical Sciences, Taipei 10617}

\author{Xiangdong Ji}
\email{xdji@sjtu.edu.cn, corresponding author}
\affiliation{Tsung-Dao Lee Institute and School of Physics and Astronomy,
Shanghai Jiao Tong Unviersity, Shanghai 200240}

\date{\today}

\begin{abstract}
Working through next-to-next-to-leading order (NNLO) accuracy in pionless effective field theory, we compute the isovector magnetic-dipole contribution to the scalar deuteron magnetic polarizability. The polarizability follows from detailed balance and a dispersion integral with the two magnetic current counter-terms fixed by the thermal $np\to d\gamma$ rate.  The conventional $\rho$ and $Z$ expansion schemes differ by $12.3\%$ at next-to-leading order but agree to $0.165\%$ at NNLO, resulting an apparent scheme dependence and supplying an independent convergence diagnostic.  We obtain $\beta_{M0}^{(M1_v)}=0.0789\pm0.0010~\mathrm{fm}^{3}$, a $1.3\%$ uncertainty, compared with $4.9\%$ previously.  
\end{abstract}

\maketitle

\emph{Introduction.---}
Quantitative theoretical predictions for strong-interaction hadronic/nuclear observables have achieved increasing precisions in recent years. The most prominent example is the strong coupling $\alpha_s(M_Z)$ which is known to better than one percent~\cite{PDG}.  Lattice calculations now reach percent-level precision for a large and ever-growing list of quantities, including single-hadron observables such as decay constants, as well as quark masses~\cite{FLAG}. However, the directly predicted hadronic observables at this precision are largely confined to the $A=0,1$ sectors and are often mixed with quantities used to set scales.  Once we consider many-nucleon/nuclear observables, percent-level predictions that experiment can verify are much more difficult and rare. 

The possibility of pursuing percent-level precision in nuclear physics emerged from a program initiated roughly thirty-five years ago.
Weinberg's proposal to organize nuclear forces by the same effective-Lagrangian logic that had already reshaped low-energy hadron physics~\cite{Weinberg1979,Weinberg1990,Weinberg1991} replaced model-dependent potentials with a systematic expansion in the ratio of a soft scale to a hard one~\JW{\cite{Beane:2000fx,Beane:2001bc,Bedaque:2002mn,Kubodera:2004zm,Meissner:2014lgi,Hammer:2012id,Epelbaum:2011md,Acharya:2023xpd}}.  In the two-nucleon sector $(A=2)$ the pionless theory was developed and carried the idea furthest, resumming the large $S$-wave scattering lengths and treating range corrections and electroweak currents perturbatively\JW{~\cite{Bedaque:1998kg,ChenRupakSavage1999,Kaplan:1996xu}}
%~\cite{KSW1998a,KSW1998b,ChenRupakSavage1999} 
after integrating out the pions in Weinberg's original proposal.  This program quickly yielded precision calculations: $pp$ fusion~\cite{ButlerChen2001} and $np\to d\gamma$ to N$^4$LO~\cite{ChenSavage1999,Rupak2000}, with the promise that few-nucleon electroweak processes could be brought under percent-level theoretical control.  Three decades of higher-order calculations and consolidation of these results for astrophysical use~\cite{SolarFusionIII2025} have gradually converted what began as an argument for systematics into a tool for precision. 

There exist now a number of precise calculations in the few-nucleon sector.  The issue is that many of these precise calculations are not experimentally verifiable.  The $pp$ fusion $S$-factor has been calculated with sub-percent theoretical uncertainty~\cite{SolarFusionIII2025} and cannot be directly measured at solar energies; its precision is an input to stellar models, but has not faced with actual data. The thermal cross section $np\to d\gamma$ has been measured at a few parts per thousand, but is used as an input to fix short-distance currents~\cite{Rupak2000}, not as a way to test them. An excellent example of experimentally-testable nuclear observables that can be predicted at the percent level includes the electric 
polarizability of the deuteron $\alpha_{E0}$~\cite{JiLi2004}. The recent high-precision experiment at Shanghai Laser Electron Gamma Source (SLEGS) measured $D(\gamma,n)p$ from $2.33$ to $19.65~\mathrm{MeV}$ obtained~\cite{Hao2026}
\begin{eqnarray}
 \alpha_{E0}+\bM
 &=& 0.719\pm0.009_{\rm stat} \nonumber \\ && \pm0.014_{\rm algo}\pm0.023_{\rm syst}~\mathrm{fm}^{3}.
 \label{eq:slegs}
\end{eqnarray}
This provides a measurement of $\alpha_{E0}$ at the $\sim 4\%$ level, assuming that $\bM$ is known to $5\%$ accuracy. Future higher precision data have the potential to make the experimental error competitive with the nuclear EFT prediction. 

Here we report a calculation of the deuteron magnetic polarizability to the precent level accuracy, to further
challenge high-precision experimental verification. This observable allows accurate theory prediction because that the isovector $M1$ transition from the bound ${}^{3}S_{1}$ channel to the near-threshold ${}^{1}S_{0}$ channel is enhanced by the large isovector nucleon magnetic moment and the large singlet scattering length~\cite{ChenEtAl1998,JiLi2004}. More importantly, the same two short-distance current counterterms that control this transition can be fixed by high-precision $np$ capture data. Future improvements of the SLEGS experiment might  provide additional important precision tests of the nuclear EFT framework.

\vspace{0.1in}
\emph{Polarizability through dispersion integral.---}
The connection between neutron capture and the magnetic polarizability follows from detailed balance and a dispersion relation.  Let $M_N$ denote the isospin-averaged nucleon mass, $B_d$ the deuteron binding energy, and
$\gamma=\sqrt{M_NB_d}$ the deuteron binding momentum.  For an $np$ pair \JW{with the nucleon momentum $p$ in the center of mass frame}, the photon energy in $np\to d\gamma$ or its inverse reaction $\gamma d\to np$ is
$\omega=(p^2+\gamma^2)/M_N$.  We denote by
$\sigma_{np}^{(M1_v)}(p)$ and $\sigma_{\gamma d}^{(M1_v)}(\omega)$
the contributions to the two cross sections from the isovector
magnetic-dipole transition, $M1_v$, which dominates the magnetic
response near deuteron breakup.  Detailed balance gives
\begin{equation}
 \sigma_{\gamma d}^{(M1_v)}(\omega)
 =\frac{2p^2}{3\omega^2}\,
 \sigma_{np}^{(M1_v)}(p).
 \label{eq:db}
\end{equation}
The $M1_v$ contribution to the static scalar magnetic
polarizability is the inverse-energy-squared moment of this
photoabsorption strength.  The Baldin sum rule~\cite{Baldin1960}
therefore gives
\begin{equation}
 \bMiv=\frac{1}{2\pi^2}
 \int_{B_d}^{\infty}
 \frac{\sigma_{\gamma d}^{(M1_v)}(\omega)}
 {\omega^2}\,\dd\omega .
 \label{eq:sumrule}
\end{equation}
which is our starting point for the calculation.

\vspace{0.1in}
\emph{Pionless-EFT amplitude and matching.---}
At momenta well below the pion mass, pions need not appear as
explicit degrees of freedom.  Their effects, together with other
short-distance physics, are encoded in local interactions, while
the low-energy two-nucleon dynamics are fixed by measured
scattering parameters.  The expansion is organized in successive
powers of a low momentum over the breakdown scale, conventionally
denoted LO, NLO, NNLO, etc.

For the transition considered here, the relevant strong-interaction
inputs are the scattering length $a_s$ and effective range $r_0$ in
the spin-singlet ${}^1S_0$ $np$ channel, and the effective range
$\rho_d$ about the deuteron pole in the spin-triplet ${}^3S_1$
channel.  We also define
$\kappa_1=(\mu_p-\mu_n)/2$, where $\mu_p$ and $\mu_n$ are the proton
and neutron magnetic moments, and dimensionless parameters
\begin{equation}
 x\equiv\gamma\rho_d,\qquad
 Z_d=\frac{1}{1-x},\qquad
 z\equiv Z_d-1,
 \label{eq:xz}
\end{equation}
together with
$\lambda\equiv-1/a_s>0$ and $D\equiv\gamma+\lambda$.
Here $Z_d$ is the residue of the two-nucleon amplitude at the
deuteron bound-state pole.  The two expansion schemes considered below differ only in how this residue is organized at finite order \JW{in the expansion}.

We use the dimensionless reduced $M1_v$ $np$ capture amplitude of
Ref.~\cite{Rupak2000},
\begin{equation*}
 \widetilde X
 =\widetilde X_0+\widetilde X_1+\widetilde X_2,
\end{equation*}
where the subscripts denote LO, NLO, and NNLO.  The leading
amplitude $\widetilde X_0$ is fixed by the one-nucleon magnetic
moment and the leading $S$-wave interactions.  Effective-range
corrections and two-nucleon magnetic currents first enter at higher
orders. In the $Z$ scheme~\cite{Phillips:1999hh},
\begin{subequations}\label{eq:zamp}
\begin{align}
 \widetilde X_1^{Z}
 &=\frac{z}{2}\widetilde X_0+C_1,
 \\
 \widetilde X_2^{Z}
 &=\left(\frac{z}{2}+C_2\right)C_1
 -\frac{z^2}{8}\widetilde X_0+C_3 .
\end{align}
\end{subequations}
Here $C_1$ contains the first singlet effective-range correction
and the NLO two-nucleon magnetic current, $C_2$ is the second
singlet-range insertion, and $C_3$ contains the NNLO two-nucleon
current.  Re-expanding the pole residue through NNLO gives the
$\rho$ scheme~\cite{ChenRupakSavage1999},
\begin{subequations}\label{eq:rhoamp}
\begin{align}
 \widetilde X_1^{\rho}
 &=\frac{x}{2}\widetilde X_0+C_1,
 \\
 \widetilde X_2^{\rho}
 &=\frac{3x^2}{8}\widetilde X_0
 +\frac{x}{2}C_1+C_2C_1+C_3 .
\end{align}
\end{subequations}
The explicit forms of $\widetilde X_0$ and $C_i$ is
given in the Supplemental Material.

The two short-distance magnetic-current coefficients entering
$C_1$ and $C_3$, denoted $\Lnp$ and $\Lt$, are determined instead from the accurately
known thermal $np\to d\gamma$ cross section.  At the thermal
relative momentum $p_0=0.003443~\mathrm{MeV}$, we impose the
order-by-order matching conditions
\begin{equation}
 \sigma^{(0)}(p_0)+\sigma^{(1)}(p_0)
 =\sigma_{\rm exp},
 \qquad
 \sigma^{(2)}(p_0)=0,
 \label{eq:matching}
\end{equation}
where $\sigma^{(n)}$ denotes the contribution to the capture cross
section first entering at order $n$, and
$\sigma_{\rm exp}=334.2~\mathrm{mb}$~\cite{Rupak2000}.
The first condition fixes the NLO current $\Lnp$, while the second
fixes the NNLO current $\Lt$ so that the measured thermal
\JW{cross section} is preserved through NNLO.  The resulting values are
\begin{align}
 (\Lnp,\Lt)_\rho
 &=(-4.5588,-1.1116)~\mathrm{fm}^{2},
 \nonumber\\
 (\Lnp,\Lt)_Z
 &=(-9.0332,\phantom{-}4.9533)~\mathrm{fm}^{2}.
 \label{eq:lecs}
\end{align}
The coefficients themselves depend on the expansion schemes of the EFT;
their different values compensate the different assignment of
deuteron-residue effects between orders.  The resulting observables
must agree up to corrections beyond the order calculated.

\vspace{0.1in}
\emph{Polarizability through NNLO.---}
With the capture amplitude fixed, Eqs.~\eqref{eq:db} and
\eqref{eq:sumrule} determine the polarizability.  Expanding
$|\widetilde X|^2$ consistently through NNLO and performing the
continuum integrals analytically gives
\begin{subequations}\label{eq:compact}
\begin{align}
 \beta_{M0,\rho}^{(M1_v)}
 &=A+\bigl[xA+B_\rho\bigr]
 +\bigl[x^2A+xB_\rho+C_\rho\bigr] \ `   ,
 \\
 \beta_{M0,Z}^{(M1_v)}
 &=A+\bigl[zA+B_Z\bigr]
 +\bigl[zB_Z+C_Z\bigr] \ , 
\end{align}
\end{subequations}
separately in two expansion schemes. 
The three terms in each line are the LO, NLO, and NNLO
contributions, respectively.  The common functions entering these
expressions are
\begin{align}
 A&=
 \frac{\alpha_{\rm em}\kappa_1^{2}}
 {6M_N\gamma^{2}}\,
 \frac{3\gamma+\lambda}{D} \ ,
 \label{eq:Aexp}\\[2pt]
 B(L)&=
 \frac{\alpha_{\rm em}\kappa_1^{2}r_0\gamma}
 {6M_ND^{2}}
 +\frac{\alpha_{\rm em}\kappa_1L}{3\pi D} \ ,
 \label{eq:Bexp}\\[2pt]
 C(L,\widetilde L)&=
 \frac{\alpha_{\rm em}\kappa_1\widetilde L}{3\pi D}
 +\frac{\alpha_{\rm em}\kappa_1^{2}r_0^{2}\gamma^{3}}
 {12M_ND^{3}}
 \nonumber\\
 &\quad+
 \frac{\alpha_{\rm em}\kappa_1r_0L\gamma^{2}}
 {6\pi D^{2}}
 +\frac{\alpha_{\rm em}M_NL^{2}\gamma}
 {12\pi^{2}D},
 \label{eq:Cexp}
\end{align}
where $\alpha_{\rm em}$ is the electromagnetic fine-structure
constant, and $L$ and $\widetilde L$ denote the two magnetic-current
coefficients.  We use
\begin{equation*}
 B_\rho=B(\Lnp^\rho),\qquad
 C_\rho=C(\Lnp^\rho,\Lt^\rho),
\end{equation*}
and analogously for the $Z$ scheme.
The LO term $A$ contains no two-nucleon magnetic-current coefficient;
all dependence on the measured thermal capture \JW{cross section} enters
at NLO and beyond.  The derivation of
Eqs.~\eqref{eq:compact}--\eqref{eq:Cexp}, including the reduction of
the continuum integrals, is given in the Supplemental
Material.

\begin{figure}[t]
\centering
\includegraphics[width=\columnwidth]{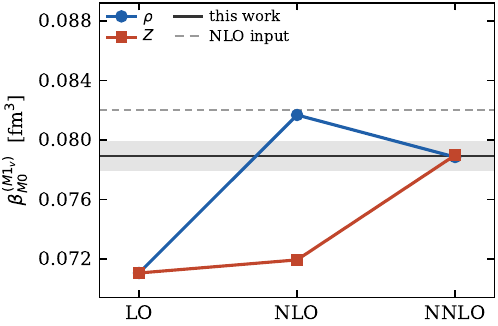}
\caption{Cumulative isovector $M1$ polarizability in the $\rho$ and
$Z$ schemes.  The two organizations differ by $12.3\%$
at NLO but agree to $1.3\times10^{-4}~\mathrm{fm}^{3}$ at NNLO.
The dashed line indicates the \JW{central value of $0.082~\mathrm{fm}^{3}$ for the} NLO input used in the SLEGS extraction~\cite{Hao2026}.
}
\label{fig:convergence}
\end{figure}

The numerical results are shown in Table~\ref{tab:orders} and
Fig.~\ref{fig:convergence}.  The two expansions give the same LO
result.  At NLO they differ by
$9.74\times10^{-3}~\mathrm{fm}^{3}$, or $12.3\%$ of the final
answer, while at NNLO they agree to
$1.30\times10^{-4}~\mathrm{fm}^{3}$, corresponding to $0.165\%$.
The recovery of this agreement at NNLO provides a nontrivial test
of the EFT expansion's convergence.

\begin{table}[b]
\caption{Order-by-order isovector $M1$ contribution to the scalar
magnetic polarizability, in $\mathrm{fm}^{3}$.  ``Shift'' denotes
the correction first entering at that order.}
\label{tab:orders}
\begin{ruledtabular}
\begin{tabular}{lrr}
 & $\rho$ scheme & $Z$ scheme \\
\colrule
LO & 0.07105 & 0.07105 \\
NLO shift & $+0.01062$ & $+0.00088$ \\
LO+NLO & 0.08167 & 0.07193 \\
NNLO shift & $-0.00283$ & $+0.00704$ \\
LO+NLO+NNLO & 0.07884 & 0.07897 \\
\end{tabular}
\end{ruledtabular}
\end{table}

The order-by-order behavior also identifies the appropriate
truncation estimate.  In the $Z$ scheme, several NLO
contributions cancel, producing an anomalously small net NLO shift~\cite{Lensky2021};
the ratio of successive corrections is therefore not representative
of the expansion.  The $\rho$ scheme instead gives
\begin{equation*}
 \left|
 \frac{\delta\beta_{M0,\mathrm{NNLO}}^{(\rho)}}
 {\delta\beta_{M0,\mathrm{NLO}}^{(\rho)}}
 \right|
 =0.266,
\end{equation*}
consistent with the nominal pionless-EFT expansion parameter
$\epsilon\equiv\gamma/m_\pi\simeq0.327$, where $m_\pi$ is the pion
mass.  We therefore estimate the first omitted correction from the
$\rho$ organization by scaling its NNLO shift,
\begin{equation}
 \Delta_{\rm EFT}
 =\epsilon
 \left|\delta\beta_{M0,\mathrm{NNLO}}^{(\rho)}\right|
 =9.3\times10^{-4}~\mathrm{fm}^{3},
 \label{eq:unc}
\end{equation}
where $\delta\beta_{M0,\mathrm{NNLO}}^{(\rho)}$ denotes the
correction first appearing at NNLO.

The remaining uncertainty from measured inputs is smaller.
Propagating the uncertainties in $\sigma_{\rm exp}$, $r_0$,
$\rho_d$, and $a_s$, while refitting the magnetic-current
coefficients in Eq.~\eqref{eq:lecs} for each variation, gives
$3.4\times10^{-4}~\mathrm{fm}^{3}$.  Combining this uncertainty in
quadrature with the EFT truncation error, taking the midpoint of the
two NNLO predictions, and rounding to the quoted precision gives
\begin{equation}
 \bMiv=0.0789\pm0.0010~\mathrm{fm}^{3}.
 \label{eq:result}
\end{equation}
The $1.3\%$ uncertainty is dominated by the estimated higher-order
EFT corrections.  The independent $\rho$ and $Z$ schemes agree
to $0.165\%$, substantially inside this uncertainty, providing an
additional convergence check on the prediction.

\vspace{0.1in}
\emph{Complete result.---}
Equation~\eqref{eq:result} gives the isovector $M1$ contribution.  The complete scalar polarizability also receives separate suppressed contributions: the known LO convection-current term, $-8.94\times10^{-4}~\mathrm{fm}^{3}$~\cite{ChenEtAl1998,JiLi2004}, and single-nucleon magnetic polarizabilities, $\beta_{M1}^{p}+\beta_{M1}^{n}=(6.2\pm1.3)\times10^{-4}~\mathrm{fm}^{3}$~\cite{PDG}, which enter as external input rather than as pionless-EFT predictions~\cite{Lensky2021}.  These two are of opposite sign and largely cancel.  For use in a Baldin-type extraction we therefore quote the nuclear combination
\begin{equation}
 \bM=0.0786\pm0.0010~\mathrm{fm}^{3},
 \label{eq:full}
\end{equation}
where the uncertainty includes an allowance for the NLO convection graphs; the Supplemental Material shows that carrying that sector beyond NLO would change Eq.~\eqref{eq:full} by less than the rounding.

Replacing the NLO input by Eq.~\eqref{eq:full} moves the electric polarizability extracted from Eq.~\eqref{eq:slegs} from $0.637$ to $0.641~\mathrm{fm}^{3}$, and reduces the theory share of the variance on that extraction from $1.9\%$ to $0.1\%$.  Because the near-threshold $M1$ strength is dominantly isovector, a multipole-separated measurement directly determines the quantity in Eq.~\eqref{eq:result}, without requiring the other sectors; the quantity in Eq.~\eqref{eq:full} is needed only when comparing against a total-photoabsorption sum rule.

\vspace{0.1in}
\emph{Possible experimental tests.---}
The prediction in Eq.~(14) provides a benchmark with $1.3\%$ uncertainty for an
experimental determination of the deuteron magnetic response.  The
best available constraint on $\beta_{M0}$, inferred from Eq.~(1), has
an uncertainty of roughly $35\%$.  A direct measurement of the
magnetic contribution would therefore provide a qualitatively new
precision test of the calculation.

The strong threshold enhancement of the $M1_v$ transition makes the
relevant experimental region particularly well defined.  Because the
moment in Eq.~(3) is weighted by $\omega^{-2}$, half of
$\beta_{M0}^{(M1_v)}$ accumulates below $2.76$ MeV, only $0.53$ MeV
above breakup, and $90\%$ below $5.43$ MeV; less than $4\%$ lies above
$10$ MeV.  The interval between threshold and the $2.33$ MeV lower
edge of Ref.~\cite{Hao2026}, only $0.11$ MeV wide, already contains
$14\%$ of the moment.  The primary experimental leverage therefore
comes from reaching as close to breakup as possible and controlling
the magnetic strength there.

The most direct test is a multipole separation of $D(\gamma,n)p$
near threshold.  Angular distributions
%, supplemented where useful by polarization observables, 
can separate the $M1$ and $E1$ strengths.
A few-percent determination of the near-threshold $M1_v$ contribution
to the weighted moment would provide a sufficient test of
Eq.~(14); a $1\%$ determination would test the prediction at its
quoted precision.

\emph{Summary.---}
We have computed the isovector $M1$ contribution to the deuteron
scalar magnetic polarizability through NNLO in pionless effective
field theory. The
short-distance magnetic currents are fixed by the thermal
$np\to d\gamma$ rate, while pionless EFT predicts the continuum
energy dependence that enters the dispersion moment.  Two
parameterizations that differ by $12.3\%$ at NLO agree to $0.165\%$
at NNLO, providing an independent convergence diagnostic and leaving
a final theoretical uncertainty of $1.3\%$.
Including the suppressed
convection-current and single-nucleon contributions, we have 
$\beta_{M0}=0.0786\pm0.0010~\mathrm{fm}^{3}$.  

Replacing the previous
NLO magnetic input by this result reduces the theory contribution to
the variance of the SLEGS electric-polarizability extraction from
$1.9\%$ to $0.1\%$.  The precision of the magnetic prediction
therefore already exceeds what is required for its present use as
theoretical input for extracting $\alpha_{E0}$.

The next step is an independent experimental test.  The
$\omega^{-2}$ weighting of the dispersion relation localizes the
required information near deuteron breakup, where the $M1_v$
transition is largest and the prediction is most directly accessible.

\section{Acknowledgments}
We thank the experimental group at SLEGS, particularly Y. G. Ma and F. G. Tao, for useful discussions and communications that motivated the present work.
R. C. acknowledges the kind hospitality of the Tsung-Dao Lee Institute in Shanghai, where this work was completed.
J.W.C. is supported by the National Science and Technology Council of Taiwan under Grants 115-2112-M-002 -016 -MY3 and 115-2918-I-002 -026.
 X.-J. is partially supported by NSFC Grant No. 12635004, the State Key Laboratory of Dark Matter Physics, and Thomas and Linda Lau Family Foundation.

\bibliography{references} 

\onecolumngrid
\section{Supplemental Material}

Supplemental Material records the conventions, NNLO capture amplitude, order-by-order matching, analytic polarizability integrals, and numerical reproduction of the result quoted in the letter.  The calculation is restricted to the  dominant isovector magnetic-dipole transition ${}^{3}S_{1}\leftrightarrow{}^{1}S_{0}$ \JW{near the threshold.}  The remaining subdominant transverse sectors are discussed separately below.

\subsection{Conventions and numerical inputs}

We use natural units, $\hbar=c=1$, and convert MeV to inverse femtometers with $\hbar c=197.3269804~\mathrm{MeV\,fm}$.  The deuteron binding momentum and the inverse singlet scattering length are
\begin{equation}
 \gamma=\sqrt{M_N B_d},
 \qquad
 \lambda\equiv-\frac{1}{a_s}>0,
 \qquad
 D\equiv\gamma+\lambda.
 \label{eq:definitions}
\end{equation}
The two organizations of the triplet effective-range expansion are described by
\begin{equation}
 x\equiv\gamma\rho_d,
 \qquad
 Z_d=\frac{1}{1-x},
 \qquad
 z\equiv Z_d-1=\frac{x}{1-x}.
 \label{eq:xz-supp}
\end{equation}
Thus $z=x+x^2+\mathcal O(x^3)$ when it is re-expanded in the conventional $\rho$ counting.  The numerical values used in the supplied NNLO calculation are listed in Table I.

\begin{table}[h]
\caption{Inputs.  Lengths are converted to inverse MeV inside the calculation and restored in the final answer.}
\label{tab:input-values}
\begin{ruledtabular}
\begin{tabular}{lll}
Quantity & Value & Comment \\
\colrule
$M_N$ & $938.92~\mathrm{MeV}$ & isospin-averaged nucleon mass \\
$B_d$ & $2.2246~\mathrm{MeV}$ & deuteron binding energy \\
$\gamma$ & $45.702~\mathrm{MeV}=0.231608~\mathrm{fm}^{-1}$ & derived \\
$\alpha_{\rm em}$ & $1/137$ & electromagnetic coupling \\
$a_s$ & $-23.75~\mathrm{fm}$ & singlet scattering length \\
$r_0$ & $2.73~\mathrm{fm}$ & singlet effective range \\
$\rho_d$ & $1.764~\mathrm{fm}$ & triplet range about the pole \\
$\kappa_1$ & $2.353$ & isovector nucleon magnetic moment \\
$p_0$ & $0.003443~\mathrm{MeV}$ & thermal capture momentum \\
$\sigma_{\rm exp}$ & $334.2~\mathrm{mb}$ & thermal $np\to d\gamma$ rate \\
\end{tabular}
\end{ruledtabular}
\end{table}

For these inputs,
\begin{equation}
 x=0.40856,
 \qquad
 z=0.69078,
 \qquad
 \frac{\gamma}{m_\pi}=0.32745
 \quad (m_\pi=139.57~\mathrm{MeV}).
 \label{eq:derivednumbers}
\end{equation}

\subsection{NNLO isovector capture amplitude}
\label{sec:amplitude}

We use the dimensionless reduced isovector $M1$ amplitude of Ref.~\cite{Rupak2000}.  In the sign convention of Eq.~\eqref{eq:definitions}, its leading term is
\begin{align}
 \widetilde X_0(p)
 &=-\kappa_1\gamma^2
 \left[
 \frac{1}{p^2+\gamma^2}
 -\frac{1}{(\gamma-ip)(-\lambda+ip)}
 \right]
 \nonumber\\
 &=\frac{\kappa_1\gamma^2D}
 {(p^2+\gamma^2)(-\lambda+ip)}.
 \label{eq:x0}
\end{align}
It is useful to define
\begin{subequations}\label{eq:c123}
\begin{align}
 C_1(p)
 &=\frac{\kappa_1 r_0 p^2\gamma^2}
 {2(-\lambda+ip)^2(\gamma-ip)}
 +\frac{\gamma^2M_N\Lnp}{4\pi(-\lambda+ip)},
 \label{eq:c1}
 \\
 C_2(p)
 &=\frac{r_0p^2}{2(-\lambda+ip)},
 \label{eq:c2}
 \\
 C_3(p)
 &=\frac{\gamma^2M_N\Lt}{4\pi(-\lambda+ip)}.
 \label{eq:c3}
\end{align}
\end{subequations}
Here $\Lnp$ and $\Lt$ are renormalization-group-invariant two-nucleon magnetic-current counterterms.  In the $Z$ parameterization the amplitude is organized as
\begin{subequations}\label{eq:zamp-supp}
\begin{align}
 \widetilde X_1^Z
 &=\frac{z}{2}\widetilde X_0+C_1,
 \\
 \widetilde X_2^Z
 &=\left(\frac{z}{2}+C_2\right)C_1
 -\frac{z^2}{8}\widetilde X_0+C_3.
\end{align}
\end{subequations}
Re-expanding $z=x+x^2+\cdots$ while keeping terms through NNLO gives the $\rho$ organization,
\begin{subequations}\label{eq:rhoamp-supp}
\begin{align}
 \widetilde X_1^\rho
 &=\frac{x}{2}\widetilde X_0+C_1,
 \\
 \widetilde X_2^\rho
 &=\frac{3x^2}{8}\widetilde X_0
 +\frac{x}{2}C_1+C_2C_1+C_3.
\end{align}
\end{subequations}
The two expressions are identical order by order after the appropriate re-expansion.  Their different numerical sequences come from keeping the pole residue $Z_d$ in different places, together with separate order-by-order fits of the current operators.

The capture cross section in this normalization is
\begin{equation}
 \sigma_{np}^{(M1_v)}(p)
 =\frac{4\pi\alpha_{\rm em}(p^2+\gamma^2)^3}
 {\gamma^3M_N^4p}\,
 \left|\widetilde X(p)\right|^2,
 \qquad
 \widetilde X=\widetilde X_0+\widetilde X_1+\widetilde X_2.
 \label{eq:capturexs}
\end{equation}
The squared amplitude is expanded as,
\begin{subequations}\label{eq:squaredexpansion}
\begin{align}
 S_0&=|\widetilde X_0|^2,
 \\
 S_1&=2\,\mathrm{Re}\!\left(\widetilde X_0\widetilde X_1^*\right),
 \\
 S_2&=|\widetilde X_1|^2
 +2\,\mathrm{Re}\!\left(\widetilde X_0\widetilde X_2^*\right).
\end{align}
\end{subequations}
Thus $|\widetilde X|^2=S_0+S_1+S_2+\mathcal O(Q^3)$.

\subsection{Order-by-order capture matching}
\label{sec:matching}

The current coefficients are fit to the known thermal capture rate. At $p=p_0$ we impose
\begin{subequations}\label{eq:matching-supp}
\begin{align}
 \sigma^{(0)}(p_0)+\sigma^{(1)}(p_0)&=\sigma_{\rm exp},
 \label{eq:matchnlo}
 \\
 \sigma^{(2)}(p_0)&=0,
 \label{eq:matchnnlo}
\end{align}
\end{subequations}
where $\sigma^{(n)}$ is obtained by replacing $|\widetilde X|^2$ in Eq.~\eqref{eq:capturexs} by $S_n$.  Equation~\eqref{eq:matchnlo} determines $\Lnp$ in each parameterization; Eq.~\eqref{eq:matchnnlo} then determines $\Lt$.  The fitted values are shown in Table II.
\begin{table}[h]
\caption{Order-by-order magnetic-current coefficients in the two parameterizations.}
\label{tab:lecs}
\begin{ruledtabular}
\begin{tabular}{lrr}
 & $\rho$ parameterization & $Z$ parameterization \\
\colrule
$\Lnp~[\mathrm{fm}^2]$ & $-4.5588$ & $-9.0332$ \\
$\Lt~[\mathrm{fm}^2]$ & $-1.1116$ & $+4.9532$ \\
\end{tabular}
\end{ruledtabular}
\end{table}

\subsection{Detailed balance and the polarizability integral}
\label{sec:sumrule}

For the inverse reaction $\gamma d\to np$, detailed balance gives
\begin{equation}
 \sigma_{\gamma d}^{(M1_v)}(\omega)
 =\frac{2M_N(\omega-B_d)}{3\omega^2}\,
 \sigma_{np}^{(M1_v)}(p)
 =\frac{2p^2}{3\omega^2}\,
 \sigma_{np}^{(M1_v)}(p),
 \qquad
 \omega=\frac{p^2+\gamma^2}{M_N}.
 \label{eq:detailedbalance-supp}
\end{equation}
The scalar magnetic polarizability is the inverse-energy-squared moment
\begin{equation}
 \bMiv
 =\frac{1}{2\pi^2}
 \int_{B_d}^{\infty}
 \frac{\sigma_{\gamma d}^{(M1_v)}(\omega)}{\omega^2}\,\dd\omega.
 \label{eq:baldin-supp}
\end{equation}
Using $\dd\omega=2p\,\dd p/M_N$ and Eq.~\eqref{eq:capturexs}, this becomes the amplitude integral
\begin{equation}
 \bMiv
 =\frac{8\alpha_{\rm em}}{3\pi M_N\gamma^3}
 \int_0^\infty
 \frac{p^2\,\dd p}{p^2+\gamma^2}
 \left|\widetilde X(p)\right|^2.
 \label{eq:masterbeta}
\end{equation}
Replacing the squared amplitude by $S_0$, $S_1$, or $S_2$ defines the LO, NLO, and NNLO contributions.

All integrations reduce to
\begin{equation}
 I_{mn}(\gamma,\lambda)
 \equiv\int_0^\infty
 \frac{p^2\,\dd p}
 {(p^2+\gamma^2)^m(p^2+\lambda^2)^n}.
 \label{eq:imn}
\end{equation}
A convenient recursive method to compute this is by noting
\begin{equation}
 I_{11}=\frac{\pi}{2(\gamma+\lambda)},
 \qquad
 I_{m+1,n}=-\frac{1}{2m\gamma}\frac{\partial I_{mn}}{\partial\gamma},
 \qquad
 I_{m,n+1}=-\frac{1}{2n\lambda}\frac{\partial I_{mn}}{\partial\lambda}.
 \label{eq:recursion}
\end{equation}
Terms without a factor $(p^2+\gamma^2)^{-1}$ also use
\begin{equation}
 I_{02}=\int_0^\infty\frac{p^2\,\dd p}{(p^2+\lambda^2)^2}
 =\frac{\pi}{4\lambda}.
 \label{eq:i02}
\end{equation}

\subsection{Closed analytic result}
\label{sec:closed}

With $D=\gamma+\lambda$, the full result through NNLO can be expressed using the three terms below  
\begin{align}
 A={}&
 \frac{\alpha_{\rm em}\kappa_1^2}{6M_N\gamma^2}
 \frac{3\gamma+\lambda}{D},
 \label{eq:A}
 \\
 B(L)={}&
 \frac{\alpha_{\rm em}\kappa_1^2r_0\gamma}{6M_ND^2}
 +\frac{\alpha_{\rm em}\kappa_1L}{3\pi D},
 \label{eq:B}
 \\
 C(L,\widetilde L)={}&
 \frac{\alpha_{\rm em}\kappa_1\widetilde L}{3\pi D}
 +\frac{\alpha_{\rm em}\kappa_1^2r_0^2\gamma^3}{12M_ND^3}
 \nonumber\\
 &+\frac{\alpha_{\rm em}\kappa_1r_0L\gamma^2}{6\pi D^2}
 +\frac{\alpha_{\rm em}M_NL^2\gamma}{12\pi^2D}.
 \label{eq:C}
\end{align}

In the $\rho$ parameterization,
\begin{subequations}\label{eq:rhoresult}
\begin{align}
 \beta_{M0,\rho}^{(0)}&=A,
 \\
 \delta\beta_{M0,\rho}^{(1)}&=xA+B(\Lnp^\rho),
 \\
 \delta\beta_{M0,\rho}^{(2)}&=x^2A+xB(\Lnp^\rho)
 +C(\Lnp^\rho,\Lt^\rho).
\end{align}
\end{subequations}
In the $Z$ parameterization,
\begin{subequations}\label{eq:zresult}
\begin{align}
 \beta_{M0,Z}^{(0)}&=A,
 \\
 \delta\beta_{M0,Z}^{(1)}&=zA+B(\Lnp^Z),
 \\
 \delta\beta_{M0,Z}^{(2)}&=zB(\Lnp^Z)
 +C(\Lnp^Z,\Lt^Z).
\end{align}
\end{subequations}

\subsection{Numerical decomposition and convergence}
\label{sec:numerics}

Table~\ref{tab:numbers} gives the separate shifts and cumulative sums.  More digits than are physically significant are retained to make the calculation reproducible.

\begin{table}[h]
\caption{Dominant isovector $M1$ contribution to the scalar magnetic polarizability, in $\mathrm{fm}^3$.}
\label{tab:numbers}
\begin{ruledtabular}
\begin{tabular}{lrr}
 & $\rho$ parameterization & $Z$ parameterization \\
\colrule
LO & $0.07105$ & $0.07105$ \\
NLO shift & $+0.01062$ & $+0.00088$ \\
LO+NLO & $0.08167$ & $0.07193$ \\
NNLO shift & $-0.00283$ & $+0.00704$ \\
LO+NLO+NNLO & $0.07884$ & $0.07897$ \\
\end{tabular}
\end{ruledtabular}
\end{table}

The midpoint and full spread are
\begin{equation}
 \overline\beta_{M0}^{(M1_v)}=0.078908~\mathrm{fm}^3,
 \qquad
 |\beta_Z-\beta_\rho|=1.2982\times10^{-4}~\mathrm{fm}^3.
 \label{eq:midpointspread}
\end{equation}
The cumulative NLO values differ by $9.7391\times10^{-3}~\mathrm{fm}^3$, $12.3\%$ of the final result; the NNLO values differ by $0.165\%$. 

\subsection{Truncation-error prescription}
\label{sec:trunc}

We base the truncation estimate on the $\rho$ organization.  Its shifts fall in the pattern
\begin{equation}
 \left|\frac{\delta\beta^{(2)}_{M0,\rho}}{\delta\beta^{(1)}_{M0,\rho}}\right|=0.266,
 \label{eq:ratio}
\end{equation}
consistent with the nominal expansion parameter $\epsilon=\gamma/m_\pi=0.32745$.  $Z$ organization has an accidentally small NLO correction, for the reason discussed in the Letter, so its order-by-order ratio is not a useful diagnostic. Scaling the last computed correction by the expansion parameter gives the first omitted term,
\begin{equation}
 \Delta_{\rm EFT}
 =\epsilon\left|\delta\beta^{(2)}_{M0,\rho}\right|
 =9.26\times10^{-4}~\mathrm{fm}^3 .
 \label{eq:trunc-supp}
\end{equation}

\subsection{Input uncertainties}
\label{sec:inputs}

The couplings $L_{np}$ and $\widetilde L_{np}$ are refitted through Eqs.~\eqref{eq:matchnlo}--\eqref{eq:matchnnlo} for every input variation, so the sensitivities below include the induced change in the matching.  Table~\ref{tab:inputs} gives the response to a $1\%$ change in each input together with an assumed input uncertainty.

\begin{table}[h]
\caption{Propagated input uncertainties, $\rho$ organization.  Each entry includes the refit of the two magnetic-current couplings.}
\label{tab:inputs}
\begin{ruledtabular}
\begin{tabular}{lrrr}
input & $\partial\beta/\partial(1\%)$ & assumed & contribution \\
 & $[\mathrm{fm}^3]$ & & $[\mathrm{fm}^3]$ \\
\colrule
$\sigma_{\rm exp}$ & $1.27\times10^{-3}$ & $0.2\%$  & $2.5\times10^{-4}$ \\
$r_0$             & $1.51\times10^{-4}$ & $1.1\%$  & $1.7\times10^{-4}$ \\
$\rho_d$          & $2.38\times10^{-4}$ & $0.5\%$  & $1.2\times10^{-4}$ \\
$a_s$             & $1.95\times10^{-3}$ & $0.05\%$ & $1.0\times10^{-4}$ \\
\colrule
total (quadrature) & & & $3.4\times10^{-4}$ \\
\end{tabular}
\end{ruledtabular}
\end{table}

The sensitivity to $\sigma_{\rm exp}$ is the largest by a wide margin: a $1\%$ shift in the thermal capture rate moves $\bMiv$ by $1.3\times10^{-3}~\mathrm{fm}^3$, comparable to the entire truncation error.  This is the expected consequence of fixing the short-distance currents to that rate, and it is the reason the precision of the present result is tied to the precision of thermal neutron capture.

Combining Eq.~\eqref{eq:trunc-supp} with the input total in quadrature gives $9.87\times10^{-4}~\mathrm{fm}^3$, and rounding the midpoint of Eq.~\eqref{eq:midpointspread} together with this uncertainty,
\begin{equation}
 \bMiv=0.0789\pm0.0010~\mathrm{fm}^3 ,
 \label{eq:quoted-supp}
\end{equation}
a $1.3\%$ determination.  The $\rho$--$Z$ spread, an order of magnitude smaller, is used only as a corroborating convergence diagnostic.

\subsection{Distribution of the sum-rule moment}
\label{sec:accumulation}

Because the moment in Eq.~\eqref{eq:baldin-supp} is weighted by $\omega^{-2}$ and the isovector $M1$ strength is enhanced immediately above breakup~\cite{Baldin1960}, the integral is concentrated very close to threshold.  Table~\ref{tab:accum} gives the fraction of $\bMiv$ accumulated below a given photon energy, evaluated with the full NNLO amplitude in the $\rho$ organization.

\begin{table}[h]
\caption{Cumulative fraction of $\bMiv$ below photon energy $\omega$.  Breakup threshold is $B_d=2.2246~\mathrm{MeV}$.}
\label{tab:accum}
\begin{ruledtabular}
\begin{tabular}{rr l}
$\omega$ [MeV] & fraction & \\
\colrule
$2.33$  & $13.9\%$ & lower edge of Ref.~\cite{Hao2026} \\
$2.50$  & $32.5\%$ & \\
$3.00$  & $60.7\%$ & \\
$4.00$  & $80.8\%$ & \\
$5.00$  & $88.2\%$ & \\
$10.00$ & $96.4\%$ & \\
$19.65$ & $98.3\%$ & upper edge of Ref.~\cite{Hao2026} \\
\end{tabular}
\end{ruledtabular}
\end{table}

Half of the moment lies below $2.76~\mathrm{MeV}$, that is, within $0.53~\mathrm{MeV}$ of threshold, and $90\%$ lies below $5.43~\mathrm{MeV}$.  Two consequences follow.  First, an experiment aiming to test Eq.~\eqref{eq:quoted-supp} directly must reach as close to breakup threshold as possible; extending the upper end of the measured range is of little value, since everything above $10~\mathrm{MeV}$ carries less than $4\%$ of the total.  Second, the interval between threshold and the $2.33~\mathrm{MeV}$ lower edge of Ref.~\cite{Hao2026}, though only $0.11~\mathrm{MeV}$ wide, contains close to $14\%$ of the isovector $M1$ moment.  The magnetic response is therefore concentrated almost entirely in the region least accessible to the existing measurement.

\subsection{Subdominant transverse sectors}
\label{sec:convection}

The leading convection-current contribution to the complete scalar magnetic polarizability is~\cite{ChenEtAl1998,JiLi2004}
\begin{equation}
 \beta_{M0}^{\mathrm{conv,LO}}
 =-\frac{\alpha_{\rm em}}{32M_N\gamma^2}
 =-8.94\times10^{-4}~\mathrm{fm}^3.
 \label{eq:conv}
\end{equation}
Adding this known LO term to the midpoint in Eq.~\eqref{eq:midpointspread} gives
\begin{equation}
 \overline\beta_{M0}^{(M1_v)}
 +\beta_{M0}^{\mathrm{conv,LO}}
 =0.0780~\mathrm{fm}^3.
 \label{eq:m1plusconv}
\end{equation}
The gauged derivative interactions and the higher-order convection graphs have not been calculated consistently, so we do not quote this as a full NNLO result.  We note here what the power counting implies about how far that calculation needs to be carried.

Equation~\eqref{eq:conv} carries no dependence on $a_s$, in contrast to the leading $M1$ term in Eq.~\eqref{eq:A}.  This is due to a selection rule, as the convection current from the nucleon kinetic term is spin independent, so it cannot drive the ${}^3S_1\to{}^1S_0$ transition that carries the magnetic-dipole response.  The singlet channel does not enter this sector.  Consequently the NLO convection calculation involves the external residue correction $\gamma\rho_d$~\cite{PhillipsRupakSavage2000}, the $C_2^{({}^3S_1)}$ range insertions, and the graphs from gauging the two-derivative contact operators, but neither $r_0$ nor $L_{np}$, and it does not inherit the large-scattering-length enhancement, so no new independently undetermined coupling is expected at this order.

With $\beta_{M0}^{\rm conv,LO}$ amounting to $1.1\%$ of the total and a nominal expansion parameter $\epsilon\simeq0.33$, successive corrections in this sector scale as $2.9\times10^{-4}$, $1.0\times10^{-4}$ and $3\times10^{-5}~\mathrm{fm}^3$.  Against the truncation error of Eq.~\eqref{eq:trunc-supp}, carrying the convection sector from LO to NLO reduces the total uncertainty by about $4\%$, and from NLO to NNLO by less than $1\%$, which gets cut off after rounding.  A consistent NLO treatment is therefore sufficient for a percent-level statement about the complete $\bM$, and is required mainly because the residue term alone contributes $\gamma\rho_d\,\beta_{M0}^{\rm conv,LO}=-3.7\times10^{-4}~\mathrm{fm}^3$.

In the transverse counting of Ref.~\cite{Lensky2021} the intrinsic nucleon magnetic polarizabilities enter at a comparable scale.  Using the current values $\beta_{M1}^{p}=(2.5\pm0.4)\times10^{-4}$ and $\beta_{M1}^{n}=(3.7\pm1.2)\times10^{-4}~\mathrm{fm}^3$~\cite{PDG}, their sum is $(6.2\pm1.3)\times10^{-4}~\mathrm{fm}^3$.
This contribution is not calculable within pionless EFT and enters as external input, with an uncertainty presently dominated by the neutron. It partially cancels Eq.~\eqref{eq:conv}: the two subdominant sectors together amount to $-2.7\times10^{-4}~\mathrm{fm}^3$, or $0.35\%$ of the total. \JW{Adding this to Eq.~(\ref{eq:quoted-supp}) yields}

%Combining Eq.~\eqref{eq:m1plusconv} with the truncation and input uncertainties of Eqs.~\eqref{eq:trunc-supp} and Table~\ref{tab:inputs}, and adding in quadrature an allowance of $2.9\times10^{-4}~\mathrm{fm}^3$ for the uncalculated NLO convection graphs, gives the nuclear combination quoted in the Letter,
\begin{equation}
 \bM=\JW{0.0786}\pm0.0010~\mathrm{fm}^3 .
 \label{eq:fullbeta-supp}
\end{equation}
which gives our final quoted result.

\end{document}